\documentclass[aps,prx,twocolumn,superscriptaddress,showpacs,floatfix,preprintnumbers,nofootinbib]{revtex4-2}

\usepackage[utf8]{inputenc}

\usepackage{cancel}

\usepackage[dvipsnames]{xcolor}      
\definecolor{lcolor}{rgb}{0.5,0,0}
\definecolor{citcolor}{rgb}{0,0.0,1}

\usepackage[breaklinks,colorlinks,urlcolor=blue,citecolor=citcolor,linkcolor=lcolor,linktoc=all]{hyperref}
\usepackage{color}
\usepackage{graphicx}	
\graphicspath{{./figures/}}
\usepackage[utf8]{inputenc}
\usepackage{amsmath} 
\usepackage{amssymb}
\usepackage[ragged]{footmisc}
\usepackage{slashed} 
\usepackage{float}
\usepackage[capitalise]{cleveref}

\makeatletter
\g@addto@macro\bfseries{\boldmath}
\makeatother

\usepackage{tikz}
\usepackage[customcolors]{hf-tikz}
\usepackage{mciteplus}

\usetikzlibrary{arrows,cd,shapes,decorations.pathmorphing,decorations.markings,shadings}
\tikzset{
  big arrow/.style={
    decoration={markings,mark=at position 1 with {\arrow[scale=4,#1]{>}}},
    postaction={decorate},
    shorten >=0.4pt},
  big arrow/.default=blue}

\newcommand{\bbone}{\text{\usefont{U}{bbold}{m}{n}1}}
\MakeRobust{\bbone}

\newcommand{\Refe}{Ref.~}
\newcommand{\Refs}{Refs.~}

\newcommand{\Lbar}{\overline{\Lambda}}

\newcommand{\muB}{\mu_\text{B}}
\newcommand{\muI}{\mu_\text{I}}

\newcommand{\nc}{N_c}

\newcommand{\nf}{N_f}

\renewcommand{\epsilon}{\varepsilon}

\newcommand{\ud}{\mathrm{d}}

\newcommand{\as}{\alpha_s}
\newcommand{\gs}{g_s}

\newcommand{\gamE}{\gamma_{\text{E}}}

\usepackage[acronym]{glossaries}
\newacronym{LO}{LO}{leading-order}

\newacronym{QM}{QM}{quark matter}

\newacronym{QCD}{QCD}{quantum chromodynamics}

\newacronym{HTL}{HTL}{Hard-Thermal-Loop}

\newacronym{UV}{UV}{ultraviolet}

\newacronym{IR}{IR}{infrared}

\newcommand{\MSbar}{$\overline{\text{MS}}$}

\newcommand{\dd}{\mathrm{d}}

\newcommand{\correct}[1]{#1}

\newcommand{\correctPRD}[1]{#1}

\begin{document}

\title{Perturbative QCD meets phase quenching: The pressure of cold quark matter}

\preprint{HIP-2024-5/TH}

\author{Pablo Navarrete}
\email{pablo.navarrete@helsinki.fi}
\affiliation{Department of Physics and Helsinki Institute of Physics, P.O.~Box 64, FI-00014 University of Helsinki, Finland}
\author{Risto Paatelainen}
\email{risto.paatelainen@helsinki.fi}
\affiliation{Department of Physics and Helsinki Institute of Physics, P.O.~Box 64, FI-00014 University of Helsinki, Finland}
\author{Kaapo Seppänen}
\email{kaapo.seppanen@helsinki.fi}
\affiliation{Department of Physics and Helsinki Institute of Physics, P.O.~Box 64, FI-00014 University of Helsinki, Finland}

\begin{abstract}
Nonperturbative inequalities constrain the \correct{thermodynamic} pressure of Quantum Chromodynamics (QCD) with its phase-quenched version, a Sign-Problem-free theory amenable to lattice treatment. In the perturbative regime with a small QCD coupling constant $\as$, one of these inequalities manifests as an $O(\as^3)$ difference between the phase-quenched and QCD pressures at large baryon chemical potential.
In this work, we generalize state-of-the-art algorithmic techniques \correct{used in collider physics} 
to address large-scale multiloop computations at finite chemical potential\correct{, by direct numerical integration of Feynman diagrams in momentum space}. Using this \correct{novel} approach, we evaluate this $O(\as^3)$ difference and show that it is a gauge-independent and small positive number compared to the known perturbative coefficients at this order. This implies that \correct{at high baryon densities}, phase-quenched lattice simulations can provide a complementary nonperturbative method for accurately determining the pressure of cold quark matter at $O(\as^3)$.
\end{abstract}

\maketitle

\section{Introduction} 

Determining the bulk thermodynamic properties of strongly interacting matter at the highest temperatures and densities reached in Nature is one of the major challenges in theoretical physics. At high temperatures and small baryon densities, experiments conducted with ultrarelativistic heavy-ion collisions (see e.g. \cite{Busza:2018rrf} for a review) and theoretical work using nonperturbative lattice simulations \cite{Guenther:2020jwe,Ratti:2018ksb} have been crucial in understanding the equilibrium properties of Quantum Chromodynamic (QCD) matter. Simultaneously to these developments, the past decade has witnessed rapid progress in the study of neutron stars (NSs). Due to the tight link between NS-matter microphysics and the macroscopic properties of the stars, NSs have become unique cosmic laboratories for studying dense QCD matter (see e.g. \cite{Baym:2017whm} for a review). What makes them particularly fascinating in the context of strongly interacting physics is that, unlike any other accessible system in the Universe, the maximal densities reached inside the cores of massive NSs may exceed the limit where nuclear matter undergoes a transition into deconfined cold quark matter (QM) \cite{Annala:2019puf,Annala:2023cwx}.

To describe the thermodynamic properties of dense QCD matter in equilibrium, one needs information about the equation of state (EOS), which relates the pressure $p$ to energy density. Ideally, the EOS should be determined using nonperturbative lattice field theory tools that have provided accurate results in collider environments. Unfortunately, the standard Monte Carlo (MC) techniques fail at finite baryon density due to the Sign Problem \cite{Hasenfratz:1983ba,Karsch:1985cb,Barbour:1986jf,deForcrand:2009zkb,Nagata:2021ugx,Brandt:2022hwy,Abbott:2024vhj}, which prevents lattice QCD simulations in the region of low temperature $T$ and large baryon chemical potential $\muB$. Consequently, the only practical methods available to reliably describe the EOS of cold and dense QCD matter are limited to Chiral Effective Theory, applicable at densities around the nuclear matter saturation density \cite{Tews:2012fj,Lynn:2015jua,Drischler:2017wtt,Drischler:2020hwi}, and weak-coupling perturbative QCD within high-density QM \cite{Kurkela:2009gj,Kurkela:2016was,Gorda:2018gpy,Gorda:2021znl,Gorda:2021kme,Gorda:2021gha,Gorda:2023mkk}. Several studies \cite{Kurkela:2014vha, Annala:2017llu,Annala:2019puf,Komoltsev:2021jzg,Somasundaram:2022ztm, Komoltsev:2023zor,Gorda:2023usm,Annala:2023cwx} have shown that systematic interpolations between low- and high-density regions, combined with information from astrophysical NS observations \cite{LIGOScientific:2017vwq, LIGOScientific:2018cki, LIGOScientific:2018hze,NANOGrav:2019jur,Nattila:2017wtj,Miller:2019cac}, can effectively constrain the behavior of the EOS across all densities, provided that accurate asymptotic limits are established at both ends. 

Additionally, further bounds for the dense QCD matter EOS can be obtained using phase-quenched (PQ) lattice QCD \cite{Kogut:2007mz}. In the phase-quenched approximation, the fermion determinant in the QCD partition function for a quark with flavor $f$, nonzero mass $m_f$, and chemical potential $\mu_f$ is replaced by its magnitude
\begin{equation}
\label{eq:fermiondet}
\begin{split}
\det(\mathcal{D}(\mu_f) ) & \to |\det(\mathcal{D}(\mu_f) )| \\
& = \sqrt{\det(\mathcal{D}(\mu_f) )\det(\mathcal{D}(-\mu_f))},
\end{split}
\end{equation}
where the Dirac operator $\mathcal{D}(\mu_f) \equiv \slashed{D} + m_f + \mu_f \gamma^0$, and the last equality follows from its $\gamma^5$-hermiticity. Here, each phase-quenched fermion corresponds to two copies of ``half-species", appearing in pairs with equal mass but opposite chemical potential. Although this formulation is unphysical, it can constrain the physically more interesting scenarios involving finite $\muB$. 

For example, consider QCD with two degenerate flavors of light up and down quarks, and a baryon chemical potential $\mu_u = \mu_d = \muB/3$. In such a scenario, it follows from \cref{eq:fermiondet} that the phase-quenched partition function is equivalent to the partition function of QCD with a nonzero isospin chemical potential $\muI$, where $\mu_u = -\mu_d = \muI/2$. Using this observation and nonperturbative QCD inequalities \cite{Vafa:1984xg}, Cohen showed in \Refe\cite{Cohen:2003ut} that the phase-quenched pressure of isospin matter $p_{\text{PQ}}$ serves as a strict upper bound on the QCD pressure of baryonic matter $p$, i.e. $p_{\text{PQ}}(3\muI/2) \geq p(\muB)$.

In NS interiors, the energy density may be sufficiently high to give rise to the onset of strange quarks, either in the form of hyperons or three-flavor QM. Chemical equilibrium under weak interactions (beta-equilibrium) with a nonzero strange quark mass requires $\mu_d = \mu_s$ and $\mu_d > \mu_u$. In this context, Fujimoto and Reddy demonstrated in \Refe\cite{Fujimoto:2023unl} that a similar bound to the two-flavor case can be derived for the pressure at nonzero $\muB$ and  $\muI$, utilizing additional nonperturbative QCD inequalities derived in \Refe\cite{Lee:2004hc}. This inequality necessitates $\muI$ as a function of $\muB$, which remains unknown unless certain model-dependent assumptions are made.

In \Refe\cite{Moore:2023glb}, Moore and Gorda recently noted that the phase-quenched formulation applies to any arbitrary linear combination of chemical potentials, not limited to $\muI$. They suggested that conducting phase-quenched lattice simulations with three-flavor QM could be the most effective method to constrain the NS EOS. Although this phase-quenched configuration of quark chemical potentials represents a completely unphysical system, it holds the potential to provide the most stringent bounds on the physical NS EOS. In particular, at high densities the relative difference between the phase-quenched pressure and the QCD pressure $p_{\text{PQ}} - p$ is perturbatively order $\as^3$ in the QCD coupling constant. In any $\mu$-region where this leading $O(\as^3)$ correction is small, it can be combined with $p_{\text{PQ}}$ to provide an improved estimate of $p$. This estimation would be perturbatively complete up to higher-order corrections in $\as$ and nonperturbative effects such as pairing gaps \cite{Son:1998uk,Malekzadeh:2006ud,Alford:2007xm,Fujimoto:2023mvc}.

In this \correct{work}, we explicitly evaluate this leading perturbative $O(\as^3)$ correction. We show that this correction, originating from a single four-loop diagram with two fermion loops connected by three gluons, is a gauge-independent and small positive number compared to the known perturbative coefficients at this order \cite{Gorda:2023mkk}. Hence, phase-quenched lattice simulations can offer a complementary nonperturbative method for accurately determining the pressure of cold and dense quark matter at $O(\as^3)$. Our diagrammatic evaluation is achieved by generalizing state-of-the-art algorithmic techniques \correct{used in collider physics} \cite{Catani:2008xa,Capatti:2019ypt,Capatti:2019edf,Capatti:2022mly}
to tackle large-scale \correct{finite density} multiloop \correct{integration directly in momentum space.}
\correct{In this approach, one first derives the three-dimensional representation of a Feynman diagram by integrating all the energy components of loop integrals via the residue theorem, followed by the systematic removal of local singularities in the remaining spatial integration space that prevents direct numerical integration, and finally performs the numerical MC evaluation of the remaining finite multi-dimensional spatial momentum integrals.} This work represents the first-ever computation of a two-particle-irreducible (2PI) four-loop diagram at finite $\mu$, paving the way toward large-scale automatization of multiloop computations in the context of field theory at finite density.

\section{Setup of the calculation}\label{sec:setup}
In the context of cold and dense QM with $\nf$ quark flavors, the pressure is given in the grand canonical ensemble by $p=-\Omega$, with Euclidean partition function
\begin{equation}
    \mathcal{Z}(\{\mu_j\}) = \int \mathcal{D}A\, e^{-S[A]} \prod^{\nf}_{j=1} \det(\mathcal{D}(\mu_j) ).
\end{equation}
The perturbative expansion of the pressure then follows as the sum of all connected vacuum Feynman diagrams. The phase-quenched version of this theory makes use of the replacement in \cref{eq:fermiondet}, which can be integrated via
\begin{equation}
\sqrt{\det{\mathcal{D}(\pm\mu_j)}} = \exp{\biggl \{\frac{1}{2}\text{tr}\,\log{\mathcal{D}(\pm\mu_j)}\biggr \}}.
\end{equation}
The corresponding Feynman rule derived from here amounts to averaging over the sign of each of the quarks' chemical potentials. To see the consequences of this averaging, consider a vacuum-type Feynman diagram $\mathcal{G}(\{\mu_j\})$ in QCD with fermion momenta $P=(p^0+i\mu_j,\vec{p})$ and bosonic momenta $Q=(q^0,\vec{q})$. Upon the replacement $\mu_j\to-\mu_j$ for all quarks in $\mathcal{G}$ (i.e. take the complex conjugate), one can make use of the integral's symmetry under flipping the signs of all the fermionic and bosonic loop momenta and the Euclidean QCD Feynman rules to obtain
\begin{equation}
    \mathcal{G}(\{-\mu_j\}) = (-1)^{v_q + v_3}\mathcal{G}(\{\mu_j\}) = (-1)^{v - v_4}\mathcal{G}(\{\mu_j\}),
\label{eq:Grel}
\end{equation}
with $v$, $v_q$, $v_3$, and $v_4$ the number of total, quark-gluon, three-gluon, and four-gluon vertices, respectively. Now, a vacuum-type graph in QCD can only have an odd number of total vertices if it contains an odd number of vertices of valency 4, i.e. four-gluon interactions (see e.g. \cite{Weinzierl:2022feynint}). Hence, the exponent $v-v_4$ in \cref{eq:Grel} is an even number, leading to $\mathcal{G}(\{-\mu_j\})=\mathcal{G}(\{\mu_j\})$. This implies two things:
\begin{enumerate}
    \item Vacuum Feynman diagrams are real numbers,
    \item Phase-quenched Feynman rules are equivalent to QCD for diagrams with a single quark loop, vanishing in the difference $p_{\text{PQ}}(\{\mu_j\})-p(\{\mu_j\})$.
\end{enumerate}
Therefore, one expects possible deviations for diagrams containing at least two quark loops. 

Let us consider a generic diagram with two quark loops carrying chemical potentials $\mu_1$ and $\mu_2$, written as
\begin{equation}
\label{eq:diagramG}
    \mathcal{G}^\xi(\mu_1,\mu_2) = \int_{Q_b} \mathcal{N}^\xi(Q_b) \Gamma^{(n)}(Q_b,\mu_1) \Gamma^{(m)}(Q_b,\mu_2),
\end{equation}
where $Q_b$ denotes the set of bosonic loop momenta running through the diagram, and $\mathcal{N}^\xi(Q_b)$ contains all the color traces and the corresponding bosonic Feynman rules, for general covariant gauges parametrized by $\xi$; $\Gamma^{(n)}$ is an insertion of a gluon $n$-point function with one quark loop (suppressing Lorentz indices and stripping off color factors). By averaging over the sign of the chemical potentials, we can compute the contribution of $\mathcal{G}$ to $p_{\text{PQ}}(\mu_1,\mu_2)-p(\mu_1,\mu_2)$ keeping in mind that vacuum diagrams are real, reducing the number of independent terms from four to only two. As reversing the sign of $\mu$ in a quark-loop insertion in \cref{eq:diagramG} is equivalent to complex conjugation, we get altogether
\begin{align}
    p_{\text{PQ}} &- p\biggr\rvert_\mathcal{G} = -\frac{1}{2} \left[ \mathcal{G}^\xi(\mu_1,\mu_2) - \mathcal{G}^\xi(\mu_1,-\mu_2) \right] \nonumber \\
    &= \int_{Q_b} \mathcal{N}^\xi(Q_b) \text{Im}\, \Gamma^{(n)}(Q_b,\mu_1) \text{Im}\, \Gamma^{(m)}(Q_b,\mu_2). \label{PQ-P_1}
\end{align}
Here, we made use of the fact that bosons do not carry chemical potentials, implying $\mathcal{N}$ is real and therefore only the imaginary parts of the quark-loop insertions \correct{give non-zero contributions to $p_\text{PQ}-p$}. 

\correct{The various cancellations occurring in \cref{PQ-P_1} can be represented pictorially as}
\begin{widetext}
\begin{equation}
\label{eq:graphG}
\begin{split}
  p_{\text{PQ}} - p\biggr\rvert_\mathcal{G} &= \biggl( \raisebox{-0.45\height}{\includegraphics[height=1.4cm]{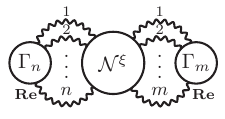}} \biggl)_\text{PQ} - \biggl( \raisebox{-0.45\height}{\includegraphics[height=1.4cm]{figures/graph_G_ReRe.pdf}} + i\times \text{(cross terms)} - \raisebox{-0.41\height}{\includegraphics[height=1.4cm]{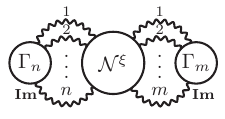}} \biggl)_\text{full QCD} \\
  &= \raisebox{-0.45\height}{\includegraphics[height=1.4cm]{figures/graph_G_ImIm.pdf}},
  \\
\end{split}
\end{equation}
\end{widetext}
\correct{where the cross terms of real and imaginary parts vanish as a consequence of vacuum diagrams being real quantities, while the effect of the subtracted phase-quenched diagram is understood as setting the contribution of the fully real combination of quark-loop insertions in full QCD to zero. Consequently, one should search for the simplest insertions of quark loops that involve non-zero imaginary parts.} One can readily see that the one-loop gluon two-point function, defined as
\begin{equation}
    \Gamma^{(2)}_{\alpha\beta}(Q,\mu_j) = \int_{\{P\}_j} \frac{\text{tr}\, [\slashed{P}\gamma_\alpha (\slashed{P}-\slashed{Q})\gamma_\beta] }{P^2(P-Q)^2},
\end{equation}
is a real \correct{function}, as complex conjugation is equivalent to the shift $P\to Q-P$, which is a symmetry of the integral. Here, \correct{$\int_{\{P\}_j}f(P^0,\vec{p})\equiv \int \frac{\ud p^0}{2\pi}\int \frac{\ud^3 \vec{p}}{(2\pi)^3} f(p^0+i\mu_j,\vec{p})$} denotes the 4-dimensional fermionic loop momentum integral.

It turns out \correct{that} this argument no longer applies to the \correct{one-loop} three-point function, as this integral has no symmetries of this type. One then expects that the leading order contribution to $p_{\text{PQ}}-p$ is a diagram composed of two quark loops connected by three gluon lines, entering at four loops: ``the Bugblatter'' \cite{Blaizot:2001vr}. In fact, there are two diagrams of this type with differing relative fermion flows, and their explicit expressions using the QCD Feynman rules, up to the terms proportional to $\xi$, are as follows:
\begin{widetext}
\begin{equation}
\label{eq:bugblatters}
\begin{split}
  \raisebox{-0.45\height}{\includegraphics[height=1.2cm]{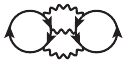}} &= \frac{\gs^6}{6}   \mathcal{T}^{abc}\mathcal{T}^{abc}
  \sum^{\nf}_{i,j=1} \biggl\{ \int_{Q \{P\}_i\{R\}^\text{*}_j S}  \frac{-\text{tr}\,[\slashed{P}\gamma^\mu (\slashed{P}-\slashed{S})\gamma^\nu (\slashed{Q}-\slashed{P})\gamma^\sigma] \, \text{tr}\,[\slashed{R}\gamma^\mu (\slashed{R}-\slashed{S})\gamma^\nu (\slashed{Q}-\slashed{R})\gamma^\sigma]}{Q^2P^2R^2S^2 (P-S)^2 (R-S)^2(Q-P)^2(Q-R)^2 (Q-S)^2} \biggl\}, \\
    \raisebox{-0.45\height}{\includegraphics[height=1.2cm]{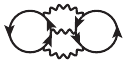}} &= \frac{\gs^6}{6} \mathcal{T}^{cba}\mathcal{T}^{abc} 
    \sum^{\nf}_{i,j=1} \biggl\{ \int_{Q \{P\}_i\{R\}_j S} \frac{+\text{tr}\,[\slashed{P}\gamma^\mu (\slashed{P}-\slashed{S})\gamma^\nu (\slashed{Q}-\slashed{P})\gamma^\sigma] \, \text{tr}\,[\slashed{R}\gamma^\sigma (\slashed{Q}-\slashed{R}) \gamma^\nu (\slashed{R}-\slashed{S}) \gamma^\mu]}{Q^2P^2R^2S^2 (P-S)^2 (R-S)^2(Q-P)^2(Q-R)^2 (Q-S)^2}
 \biggl\}. \\
\end{split}
\end{equation}
\end{widetext}
Here, $\gs = \sqrt{4\pi\as}$, and we have made use of the shorthand notation $\mathcal{T}^{abc} \equiv \text{tr}\,(T^a T^b T^c)$ for the color trace in quark loops. Note the $\mathcal{T}^{cba}$ factor in the second line of \cref{eq:bugblatters}, coming from the relative flow orientation of the $\nf$ quark flavors running in the loops, which we sum over.

Denoting the integral expressions inside the curly brackets for each diagram by $B^\xi_1(\mu_i,\mu_j)$ and $B^\xi_2(\mu_i,\mu_j)$, respectively, the Dirac traces can be related through charge conjugation, giving
\begin{equation}
    B^\xi_2(\mu_i,\mu_j) = - B^\xi_1(\mu_i,-\mu_j). \label{Furry}
\end{equation}
This is just a manifestation of Furry's theorem, as pointed out in \Refe\cite{Moore:2023glb}. On the other hand, in terms of the $\text{SU}(\nc)$ structure constants $f^{abc}$ and the totally symmetric group invariant $d^{abc}$, the color traces entering \cref{eq:bugblatters} read
\begin{equation}
\label{eq:colortraces}
    \mathcal{T}^{abc} = \frac{1}{4} (d^{abc}+if^{abc}) = (\mathcal{T}^{cba})^\text{*},
\end{equation}
implying that the Bugblatter diagrams contain the two color structures $d^{abc}d^{abc}$ and $f^{abc}f^{abc}$. Crucially, the complex conjugation property in \cref{eq:colortraces} results in a relative sign in the $f^{abc}f^{abc}$ pieces. 

With these relations at hand, we proceed to construct $p_{\text{PQ}}-p$ as shown above, by averaging over the $\nf$ chemical potentials running in the loops, for both diagrams. As instructed in the NS setup, we specialize to the case of asymptotically dense three-flavor $(\nf = 3)$ unpaired QM in beta-equilibrium, where quarks can be considered massless and quark chemical potentials are equal. We set $\mu_i=\muB/3$ for all quarks, obtaining altogether 
\begin{equation}
\begin{split}
    p_{\text{PQ}}(\muB)&-p(\muB)=\frac{-3^2}{6\times 3^4} \frac{d^{abc}d^{abc}}{16} \gs^6 \\
    &\times \left[ B^\xi_1(\muB,\muB)-B^\xi_1(\muB,-\muB) \right] + O(\gs^8), \label{PQ-P_2}
\end{split}
\end{equation}
where an additional factor of $3^4$ appears through the scaling of $B^\xi_1$ upon replacing $\mu_i^4$ by $\muB^4$. Note that the $f^{abc}f^{abc}$ part \correct{cancels} in the difference $p_{\text{PQ}}-p$, and therefore only $d^{abc}d^{abc} = (\nc^2-1)(\nc^2-4)/\nc$ survives. \correct{Notably, the resulting combination of diagrams entering \cref{PQ-P_2} is precisely of the form found in the Bugblatter contribution to QED, with the vanishing $f^{abc}f^{abc}$ piece being of purely non-Abelian nature. As a result,} the expression in \cref{PQ-P_2} satisfies several properties, as we demonstrate in the following. 

Firstly, expanding the diagrams in general covariant gauge in terms of scalarized integrals, one obtains a quadratic polynomial in the gauge parameter $\xi$. At this point, one can already appreciate the complexity of these diagrams, as they contain all nine non-factorized planar QCD vacuum topologies at four loops \cite{Navarrete:2022adz}. Upon exploiting linear relations among these integrals, in the form of momentum shifts stemming from the graphs' internal symmetries \cite{Navarrete:2024ruu, Karkkainen:2024prep}, we explicitly checked that the gauge parameter cancels out at the integrand level in the sum entering \cref{PQ-P_2}, rendering this combination gauge invariant. Consequently, in the following, we set $\xi = 0$ and drop the superscript $\xi$. 

Secondly, we can prove that this expression is automatically positive, as expected from the nonperturbative inequality $p_{\text{PQ}}\geq p$. In effect, the prefactor in \cref{PQ-P_2} is nonpositive for all $\nc$, so it suffices to show that the difference inside the square brackets is negative, a combination being precisely of the form encountered in \cref{PQ-P_1}. As bosonic propagators are real and positive in the whole region of integration, one has
\begin{equation}
\begin{split}
    B_1(\muB,\muB) & - B_1(\muB,-\muB) \\
    & = -2 \int_{QS} \frac{ \left[ \text{Im}\, \Gamma^{(3)}_{\mu\nu\sigma}(Q,S,\muB) \right]^2}{Q^2 S^2 (Q-S)^2} \label{b1}
     \leq 0,
\end{split}
\end{equation}
proving that the leading-order contribution to $p_\text{PQ}-p$ is positive.

Thirdly, we can argue that it is infrared (IR) finite. Indeed, the infrared physics of these diagrams is properly captured by the hard-thermal-loop (HTL) limit of the three-gluon vertex function, whose color structure is $\Gamma^{(3)}_\text{HTL} \sim f^{abc}$ \cite{Gorda:2021kme}. This means that only the $f^{abc}f^{abc}$ piece of the Bugblatter diagrams, canceling in the difference $p_\text{PQ}-p$, develops a logarithmic infrared divergence. Consequently, the structure $d^{abc}d^{abc}$ in \cref{PQ-P_2} which we compute here must be infrared finite (see also \cite{Moore:2023glb}), further implying that non-analytic logarithms in the coupling arising from the HTL resummation do not appear at this order, but can potentially enter only at order $O(\as^4\ln^n\as)$.

Finally, we argue that \cref{b1} is ultraviolet (UV) finite. One might expect this based on the observation that there are no lower loop diagrams with the color factor $d^{abc}d^{abc}$ that could renormalize \cref{PQ-P_2}. Indeed, this combination picks up the imaginary part of the vertex function, a quantity that vanishes in the ultraviolet due to this limit being pure vacuum (i.e. $i \mu\to 0$). Thus, only the $f^{abc}f^{abc}$ piece of the Bugblatter diagrams entering full QCD, containing the real part of the vertex function, requires renormalization.

\section{A new numerical approach}

\correct{To proceed with the evaluation of the fully finite four-loop momentum integrals displayed in \cref{b1}, we next introduce a novel approach to numerical multiloop integration at finite density.}

In modern quantum field theory, multiloop \correct{Feynman} diagrams are often first expanded in terms of scalarized integrals, which are then systematically reduced to a basis of so-called master integrals through large-scale automated algorithmic procedures \correct{based on solving integration-by-parts (IBP) identities} \cite{Chetyrkin:1981qh,Laporta:2000dsw}. In this context, techniques for computing master integrals in dimensional regularization are highly developed \cite{Henn:2013pwa}. However, in a thermal setup (finite $\mu$ and/or $T$) we break Lorentz symmetry, limiting the usage of the traditional techniques that have proven successful in the vacuum setting. \correct{Despite this, all thermal cases relevant to the pressure up to the three-loop level have been managed on a case-by-case basis \cite{Vuorinen:2003fs,Schroder:2012hm}. This is achieved through brute-force semi-analytical calculation of the scalarized integrals, which rely on tedious subtractions of divergences at the level of one-loop self-energy insertions}. \correct{At the four-loop level, the presence of integral topologies containing more complicated substructures severely restricts the implementation of this method to only the simplest cases \cite{Gynther:2007bw,Gynther:2009qf,Gorda:2022zyc}, calling for the development of new tools and methods to handle multiloop calculations beyond three loops.}

\correct{At finite density, proposed approaches in this direction include the so-called cutting rules method, which was introduced in Refs. \cite{Kurkela:2009gj,Ghisoiu:2016swa}, and a generalization of IBP to finite density, developed in \Refe\cite{Osterman:2023tnt}. While the latter has not yet been explored even at three loops, the former has been successfully employed at this level in \Refe\cite{Kurkela:2009gj}.  In the cutting rules, a representation of Feynman integrals is derived by integrating the energy components of fermionic loops via the residue theorem, resulting in a sum of three-dimensional phase-space integrals over on-shell amplitudes at zero $\mu$. However, a generic obstacle in this approach is that the dimensionally regulated phase-space integrals introduce spurious IR divergences not present in the original Feynman integral, complicating their calculation even for the simplest four-loop topologies.}

Recent developments in high-performance numerical evaluation of \correct{Feynman diagrams} directly in momentum space, such as the Loop-Tree Duality (LTD) (see e.g. \cite{Catani:2008xa,Capatti:2019ypt}), offer a promising alternative avenue to multiloop integration. The defining feature these techniques capitalize on is the analytical computation of \correct{both fermionic and bosonic energy} integrals via the residue theorem, \correct{in contrast to the cutting rules method}. This approach is particularly suitable for finite-density calculations due to the effect of the chemical potentials being entirely encoded in the temporal momentum components.

Inspired by the LTD and related ideas, we put forward a novel generalization of these approaches to finite-density field theory, which we call dense LTD (dLTD). Specifically, we implement numerical MC integration routines for the spatial momentum integrals, preceded by an analytical computation of zero-component \correct{(energy)} residues. This is realized by restructuring the integrand into a convenient representation, allowing for the \correct{systematic} removal of local spurious (integrable) singularities that hinder direct numerical integration.

\subsection{Derivation of dLTD}

In the following, we provide a finite-$\mu$ generalization of the algorithm introduced in \Refe\cite{Capatti:2022mly} for computing the \correct{energy} integrals, employing the notation introduced therein. Let $\mathbf{x}=(x_i)_{i\in\{1,\dots,n\}}$ denote a vector with $n$ components, $\mathbf{x}\cdot\mathbf{y} = \sum_{i=1}^nx_iy_i$ the standard scalar product of two vectors $\mathbf{x}$ and $\mathbf{y}$, and $\mathbf{x}\odot\mathbf{y} = (x_iy_i)_{i\in\{1,\dots,n\}}$ component-wise multiplication of the vectors. Now, a vacuum-type $l$-loop integral with $n$ single-power propagators at finite $\mu$, stripped of its spatial loop integrations over $\vec{p}_i$, can be written as
\begin{equation}
    I \equiv \int\left[ \prod_{i=1}^l\frac{\ud p_i^0}{2\pi}\right]\frac{\mathcal{N}(\mathbf{Q}^0)}{\prod_{j=1}^n\left[\left(Q_j^0\right)^2+E_j^2\right]},\label{eq:Iintegral}
\end{equation}
where the numerator $\mathcal{N}$ is a regular function of the \correct{propagator energies} $\mathbf{Q}^0=\sum_{i=1}^l\mathbf{S}_ip_i^0+i\boldsymbol{\varphi}\mu$, the vectors $\mathbf{S}_i = (S_{1i},\dots,S_{ni})$ with $S_{ji}\in\{\pm 1,0\}$ fix the loop momentum basis, and the vector $\boldsymbol{\varphi} = \sum_{i=1}^l\mathbf{S}_i\tilde{\varphi}_i\in\{\pm 1,0\}^n$ with $\tilde{\varphi}_i\in\mathbb{Z}$ determines the fermion signature of each propagator. The propagator \correct{on-shell} energies are given by $E_j^2 = |\vec{q}_j|^2+m_j^2$, where the spatial loop momenta are $\vec{\mathbf{q}} = \sum_{i=1}^l\mathbf{S}_i\vec{p}_i$ and the masses are denoted by $m_j$.

\correct{
We start the derivation by inserting the unity
\begin{equation}
    1=\int\dd\tilde{p}_j^0\delta\left(\tilde{p}_j^0-\sum_{i=1}^lS_{ji}p^0_i\right)
\end{equation}
for each propagator in \cref{eq:Iintegral} and writing the Dirac deltas as integrals over the variables $\tau_j$,
\begin{equation}
\begin{split}
    &I = \int\prod_{i=1}^l\frac{\dd p_i^0}{2\pi}\int\prod_{j=1}^n\dd\tau_j\,\mathrm{e}^{-i\tau_j\sum_{i=1}^lS_{ji}p^0_i} \\
    &\times\int\left[\prod_{j=1}^n\frac{\dd\tilde{p}_j^0}{2\pi}\frac{\mathrm{e}^{i\tau_j\tilde{p}_j^0}}{\left(\tilde{p}_j^0+i\varphi_j\mu\right)^2+E_j^2}\right]\mathcal{N}\left(\tilde{\mathbf{p}}^0+i\boldsymbol{\varphi}\mu\right).\label{eq:Iintegral1}
\end{split}
\end{equation}
Now the energy integrals over $\tilde{p}_i^0$ can be performed for each propagator independently. Using residue calculus, we obtain
\begin{equation}
\begin{split}
    I &= \int\prod_{i=1}^l\frac{\dd p_i^0}{2\pi}\int\prod_{j=1}^n\dd\tau_j\,\mathrm{e}^{-i\tau_j\sum_{i=1}^lS_{ji}p^0_i} \\
    &\times\Biggl[\prod_{j=1}^n\frac{1}{2E_j}\sum_{\rho_j,\sigma_j\in\{\pm 1\}}\rho_j\sigma_j \mathrm{e}^{-\tau_j(\rho_j E_j-\varphi_j\mu)} \\
    &\times\Theta\bigl(\sigma_j(\rho_j E_j-\varphi_j\mu)\bigr)\Theta(\sigma_j\tau_j)\Biggr]\mathcal{N}(i\boldsymbol{\rho}\odot\mathbf{E}).
\end{split}
\end{equation}
Changing the integration variables as $\boldsymbol{\tau}\to\boldsymbol{\sigma}\odot\boldsymbol{\tau}$, renaming $\boldsymbol{\rho}\odot\boldsymbol{\sigma}\to\boldsymbol{\rho}$, $\boldsymbol{\sigma}\to\boldsymbol{\sigma}$, writing the integrals over $p_i^0$ as Dirac delta functions, and reorganizing yields
\begin{equation}
\begin{split}
    I &= \sum_{\substack{\boldsymbol{\rho}\in\{\pm 1\}^n \\ \boldsymbol{\sigma}\in\{\pm 1\}^n}}\mathcal{N}(i\boldsymbol{\sigma}\odot\boldsymbol{\rho}\odot\mathbf{E})\prod_{j=1}^n\frac{\Theta(\rho_jE_j-\sigma_j\varphi_j\mu)}{2\rho_jE_j} \\
    &\quad\times\int_{\mathbb{R}_+^n}\mathrm{d}\boldsymbol{\tau}\,\mathrm{e}^{-\boldsymbol{\tau}\cdot(\boldsymbol{\rho}\odot\mathbf{E})}\prod_{i=1}^l\delta\bigl(\boldsymbol{\tau}\cdot(\boldsymbol{\sigma}\odot\mathbf{S}_i)\bigr),\label{eq:Iresult}
\end{split}
\end{equation}
where the exponential has simplified due to the property $\boldsymbol{\tau}\cdot(\boldsymbol{\sigma}\odot\boldsymbol{\varphi})=\sum_{i=1}^l\tilde{\varphi}_i \boldsymbol{\tau}\cdot(\boldsymbol{\sigma}\odot\mathbf{S}_i)$. Before this simplification, the expression in the parenthesis in the exponent matches the arguments of the step functions. 
}

In contrast to the vacuum case \correct{presented in \Refe\cite{Capatti:2022mly}}, the generalization in \cref{eq:Iresult} involves step functions that include $\mu$, along with an additional sum over the sign vectors $\boldsymbol{\rho}$. However, the integral on the second line of \cref{eq:Iresult}, representing a Laplace transform of a non-simplicial convex cone, is an object already encountered in the vacuum computation. We perform this integral using the algorithmic procedure described in \Refe\cite{Capatti:2022mly}, leading to a sum of rational functions of linear combinations of the \correct{on-shell} energies $E_i$. The resulting representation for the spatial integral is free of spurious divergences, making it particularly suitable for direct numerical integration.

Nonetheless, the expression resulting from \cref{eq:Iresult} often contains integrable singularities that can impair the accuracy of MC integration. To address this issue, we have implemented a multichanneling technique \correct{similar to the one presented in \Refe\cite{Capatti:2019edf}, as detailed in \cref{sec:multichanneling}. This procedure decomposes the integrand into a sum of multiple terms (channels), with each channel parametrized in such a way that its Jacobian flattens integrable singularities specific to that channel.}

\subsection{Cross-checks of dLTD}

To verify the validity of our numerical approach, we have cross-checked that our dLTD implementation in \cref{eq:Iresult} reproduces several analytically known cases in both vacuum and finite-density field theory. As a baseline check, we first compute a nontrivial UV- and IR-finite, fully massive four-loop Bugblatter-type diagram with a trivial numerator at zero density. The comparison between our numerical result using the zero-density limit of dLTD and the analytical result from \Refe\cite{Laporta:2002pg} is displayed in the first row of \cref{eq:table}. \correctPRD{Note that our definition of the integral differs from the one used in \Refe\cite{Laporta:2002pg} only by an overall factor of $(4\pi)^{-8}$.} Using the \textsc{Vegas5.6} integration routine \cite{vegasgit} with approximately $10^9$ MC samples, we obtained $2.16931(4)\cdot 10^{-9}$, which matches the analytical result $2.16928\cdot 10^{-9}$ to the first five digits. For the finite density case, we perform comparisons between our dLTD results and analytical calculations for UV-subtracted, IR-finite, two- and three-loop vacuum-type QCD diagrams \correctPRD{in Feynman gauge ($\xi=0$)} with massless quarks \cite{Vuorinen:2003fs}. Details for these cases can be found in \cref{sec:twoloopexample}. The results are shown in the last two rows of \cref{eq:table}, where we observe an excellent agreement including the first four digits. In the last column of the table, we present the \correctPRD{time required to evaluate one sample on a single core}, demonstrating efficient scaling on the number of loops. \correctPRD{Here, $N$ represents the total number of samples used to evaluate a single diagram.}

\begin{table}
\centering
\begin{tabular}{||c|c|c|c|c|c|}
    \hline
    Diagram & Analytic & dLTD & $N\,[10^6]$ & $[\mathrm{\mu s}]$ \\
    \hline \hline
    \raisebox{-0.55\height}{\includegraphics[height=0.85cm]{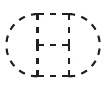}} &
    \raisebox{-0.12cm}{$2.16928\cdot 10^{-9}$ \cite{Laporta:2002pg}} & \raisebox{-0.12cm}{$2.16931(4)\cdot 10^{-9}$} & \raisebox{-0.12cm}{3000} & \raisebox{-0.12cm}{5.2} \\
    \noalign{\hrule height.9pt} \noalign{\hrule height.9pt}
     \raisebox{-0.55\height}{\includegraphics[height=0.85cm]{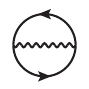}} & \raisebox{-0.12cm}{$-0.00128325$ \cite{Vuorinen:2003fs}} & \raisebox{-0.12cm}{$-0.00128338(23)$} & \raisebox{-0.12cm}{150} & \raisebox{-0.12cm}{5.8} \\
     \hline
     \raisebox{-0.55\height}{\includegraphics[height=0.85cm]{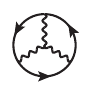}} & \raisebox{-0.12cm}{$-0.000256547$ \cite{Vuorinen:2003fs}} & \raisebox{-0.12cm}{$-0.00025654(9)$} & \raisebox{-0.12cm}{800} & \raisebox{-0.12cm}{9.6} \\
     \hline
\end{tabular}
\caption{Numerical integration via our approach compared against analytic results from \Refs\cite{Laporta:2002pg,Vuorinen:2003fs} for IR- and UV-finite massive four-loop scalar diagram at $\mu=0$ and IR-finite two- and three-loop QCD diagrams at finite $\mu$ (here set to 1). The UV divergence of the three-loop diagram has been subtracted away in the \MSbar~scheme, while $\gs=\nf=1$, $\nc=3$, and $\Lbar=\mu$. $N$ denotes the MC statistics and $[\mathrm{\mu s}]$ is the \correctPRD{time needed to evaluate one sample on a single core} in microseconds.}
\label{eq:table}
\end{table}

\begin{figure}
    \centering \includegraphics[width=0.43\textwidth]{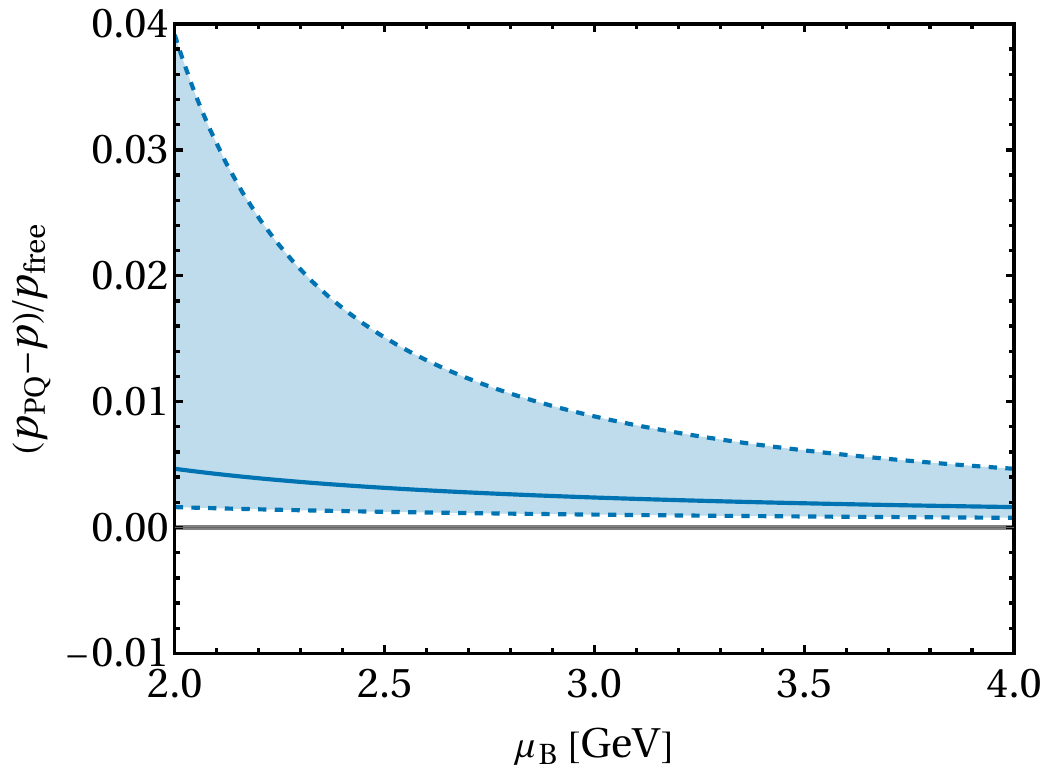}
    \caption{The leading-order difference between the pressure of three-flavor dense QM and its phase-quenched version, normalized by the free pressure, as a function of the baryon chemical potential. The shaded region shows the scale variation of $X\in[1/2,2]$, while the solid curve is the central value $X=1$.}
    \label{fig:pPQ-p}
\end{figure}

\section{Results and discussion}

We are now prepared to evaluate the new coefficient in \cref{b1} using the integrand representation generated by \cref{eq:Iresult}. With a suitable choice of coordinates \correct{(see \cref{sec:parametrization})}, this leads to a nine-dimensional integral over the spatial momenta. We compute this integral numerically employing the \textsc{Vegas5.6} integration routine \cite{vegasgit}, distributing roughly $6\cdot 10^{11}$ MC samples across 75 channels (see \cref{sec:multichanneling}) and spending 7 microseconds per sample \correctPRD{on a single core} on average. We determine the new coefficient as
\begin{equation}
B_1(\muB,\muB) - B_1(\muB,-\muB) = -3.327(7)\cdot 10^{-6}\muB^4,   
\end{equation}
where the MC error in the numerical result is \correct{negligible for practical purposes.  Additionally, the cancellation of IR divergences in \cref{b1} is non-local, necessitating a regularization procedure for numerical evaluation. To address this, we introduce a small but nonzero mass in the gluon propagators, which effectively screens these residual IR singularities. Since the integrated result is IR finite, as discussed in \cref{sec:setup}, it should remain insensitive to the small value of the regulator. We have verified this insensitivity by varying the gluon mass within the range $[10^{-5}\muB,10^{-2}\muB]$ (see \cref{sec:gluonmass}).}

By utilizing \cref{PQ-P_2} and setting $\nc=3$, we then find that the relative difference $p_{\text{PQ}} - p$ in the perturbative region is
\begin{equation}
    p_{\text{PQ}}(\muB) - p(\muB) = 3.368(7)\cdot\left (\frac{\as(\Lbar)}{\pi}\right )^3 p_{\text{free}},
\end{equation}
where $\as = \as(\Lbar)$ represents the renormalized strong coupling constant in the \MSbar~scheme at the renormalization scale $\Lbar$, and we have written our result in terms of the pressure of a free Fermi gas of quarks in beta equilibrium, denoted as $p_{\text{free}} = 3(\muB/3)^4/4\pi^2$. 

In \cref{fig:pPQ-p}, we display the relative difference $p_{\text{PQ}}- p$ normalized by the free quark pressure as a function of $\muB$. We employ the known three-loop running for $\as(\Lbar)$, set $\Lbar = 2X\muB/3$, and estimate the uncertainty arising from truncating the perturbative series by varying $X$ in the range $[1/2, 2]$. It is observed that in the perturbative region $\muB \gtrsim 2.5$ GeV, corresponding to $\as \lesssim 0.5$, the difference $p_{\text{PQ}} - p$ is very small. This implies that phase-quenched lattice simulations provide an accurate approximation for the dense QM pressure in the perturbative regime.

A few comments on the significance of our results are in order. First, we have \correct{generalized state-of-the-art algorithmic techniques used in collider physics to handle large-scale multiloop computations at finite chemical potential, by direct numerical integration of Feynman diagrams in momentum space. This novel technique allowed us to perform the first-ever finite-$\mu$ computation of a 2PI four-loop diagram, being part of the most complicated vacuum topology in QCD.} \correct{We stress that our new approach, which we call dLTD, is not limited to the calculation presented in this work, but it is equally applicable to diagrams of various topologies at the four-loop level, in addition to holding the prospect of being generalizable to the case of diagrams with external legs}. Second, our numerical evaluation of the leading $O(\as^3)$ perturbative correction to the difference $p_{\text{PQ}}(\muB) - p(\muB)$ demonstrates that Sign-Problem-free lattice simulations can provide a complementary method for accurately evaluating the pressure of cold and dense QM at $O(\as^3)$. Finally, we present an estimate for the necessary accuracy of phase-quenched lattice simulations to effectively constrain the EOS of QM at high density. Drawing upon the latest state-of-the-art perturbative findings (see the right panel in Fig.~2 of \Refe\cite{Gorda:2023mkk}), we approximate that the uncertainty in phase-quenched lattice simulations should be less than 25\% for $\muB \geq 2.2$ GeV.

As a closing remark, we believe the results presented in this paper will play a significant role in achieving the goal of determining the $O(\as^3)$ EOS of dense quark matter. We emphasize that this could have a major impact on the inference of the neutron star matter EOS, potentially leading to a true breakthrough in understanding the properties of matter in neutron-star cores.

\begin{acknowledgments}
We thank Zeno Capatti, Valentin Hirschi, Aleksi Kurkela, Heikki Mäntysaari, York Schröder, Saga Säppi, and Aleksi Vuorinen for their helpful comments and suggestions. \correctPRD{Lastly, we wish to thank the anonymous referee for their insightful comments which led to an improvement in several parts of the article.} P.N., R.P., and K.S. have been supported by the Academy of Finland grants no. 347499 and 353772. P.N. is also supported by the Doctoral School of the University of Helsinki, and K.S. is additionally supported by the Finnish Cultural Foundation. We also wish to acknowledge CSC -- IT Center for Science, Finland, for computational resources. Feynman diagrams were drawn using Axodraw \cite{Vermaseren:1994je}.
\end{acknowledgments}

\appendix

\vspace{0.5cm}

\section{Multichanneling at finite density}
\label{sec:multichanneling}

Loop momentum integrals, after integrating over the temporal direction (see e.g. \cref{eq:sunset3d}), often exhibit sharp enhancements, which become integrable singularities in the case of massless propagators. These singularities can significantly hinder numerical integration. However, by employing a technique called multichanneling, it is possible to systematically remove these enhancements, provided their location and approximate functional form are known. We have implemented the multichanneling method, based on the approach described in \Refe\cite{Capatti:2019edf}, and modified it to accommodate the singularity structure specific to finite-$\mu$ loop integrals.

At vanishing $\mu$, the general functional form of the enhancements is given by \cite{Capatti:2019edf}:
\begin{equation}
    \sum_{b\in\mathcal{B}} \prod_{i\in b} E_i^{-1},\label{eq:multichanformvacuum}
\end{equation}
where $\mathcal{B}$ denotes the set of all loop momentum bases, and $E_i$ is the on-shell energy of propagator $i$. At finite $\mu$, the form of the bosonic enhancements remains unchanged, as bosons do not carry $\mu$. However, fermionic enhancements at $E_i=0$ cancel when $\mu>0$. This is evident from \cref{eq:Iresult}; in the limit where a fermionic on-shell energy vanishes, the $E_i^{-1}$ enhancement cancels in the sum over $\boldsymbol{\rho}$. Instead, fermionic singularities can occur at the intersections of Fermi surfaces, $E_i=E_j=\mu$, where the argument of the exponent in \cref{eq:Iresult} can vanish. After performing the $\boldsymbol{\tau}$ integral, these integrable singularities take the form
\begin{equation}
    \frac{\Theta(E_i-\mu)\Theta(\mu-E_j)}{E_i-E_j},
\end{equation}
where the on-shell energies $E_i$ and $E_j$ approach $\mu$ from different directions. Therefore, a functional form that captures both the bosonic and at least part of the fermionic enhancements at finite $\mu$ is
\begin{equation}
    \sum_{b\in\mathcal{B}} \prod_{i\in b} \bigl|E_i-|\varphi_i|\mu\bigr|^{-1},\label{eq:multichanform}
\end{equation}
where $\varphi_i$ is the fermion signature of propagator $i$.

Let $\mathcal{I}$ be an integrand of a spatial loop integral. To flatten the above enhancements in $\mathcal{I}$, we divide and multiply $\mathcal{I}$ by \cref{eq:multichanform} and then split up the sum over $\mathcal{B}$ in the numerator into $|\mathcal{B}|$ channels,
\begin{equation}
\begin{split}
    \mathcal{I} &= \sum_{b\in\mathcal{B}}\mathcal{C}_b, \\
    \mathcal{C}_b &\equiv \underbrace{\frac{\mathcal{I}}{\sum_{c\in\mathcal{B}} \prod_{j\in c} \bigl|E_j-|\varphi_j|\mu\bigr|^{-1}}}_{\text{enhancements cancel}}\prod_{i\in b} \bigl|E_i-|\varphi_i|\mu\bigr|^{-1}.
\end{split}
\end{equation}
The enhancements canceled by the factor in \cref{eq:multichanform} are now split into different channels specified by the loop momentum bases $b$. To remove these enhancements, we parameterize each channel $\mathcal{C}_b$ differently. We choose the loop momentum variables to be the momenta of the propagators in $b$, $\{\vec{p}_i\}_{i\in b}$, so that the Jacobian introduced in \cref{sec:parametrization} flattens the peak structure $\prod_{i\in b}|E_i-|\varphi_i|\mu|^{-1}$.

\section{Parametrization for the spatial integrals}
\label{sec:parametrization}

The Monte-Carlo integrator \textsc{Vegas5.6} \cite{vegasgit} is employed to compute the spatial integrals over $\mathbb{R}^{3l}$, where $l$ represents the number of loops. This integrator generates points within the unit hypercube $[0,1]^{3l}$. To create sample configurations for the spatial loop momenta, these points are then transformed to $\mathbb{R}^{3l}$ using a Cartesian product of spherical parametrizations \cite{Capatti:2019edf}. Specifically, for each loop momentum $\vec{p} = (p_x, p_y, p_z)$, a point in the 3D unit cube, $(x_1, x_2, x_3) \in [0,1]^3$, is mapped to $(p_x, p_y, p_z) \in \mathbb{R}^3$ using the transformation
\begin{align}
p_x &= r\sin\theta\cos\phi,  &  r &= h_a(x_1),  \nonumber \\
p_y &= r\sin\theta\sin\phi,  &  \theta &= \arccos(-1+2x_2),  \label{eq:param}\\
p_z &= r\cos\theta,   &   \phi &= 2\pi x_3, \nonumber
\end{align}
with
\begin{equation}
h_a(x) = \begin{cases} 
      p_\mathrm{F}-\left|\frac{\mu^ax}{1-x}-p_\mathrm{F}^a\right|^{1/a}, & x < \frac{p_\mathrm{F}^a}{\mu^a+p_\mathrm{F}^a} \\
      p_\mathrm{F}+\left|\frac{\mu^ax}{1-x}-p_\mathrm{F}^a\right|^{1/a}, & x \geq \frac{p_\mathrm{F}^a}{\mu^a+p_\mathrm{F}^a} 
   \end{cases},
\end{equation}
and the Jacobian
\begin{equation}
J = \frac{4\pi}{a\mu^a} r^2 |r-p_\mathrm{F}|^{1-a} \left(\mathrm{sgn}(r - p_\mathrm{F}) |r-p_\mathrm{F}|^a+p_\mathrm{F}^a+\mu^a\right)^2.
\end{equation}
For fermionic loop momenta $\vec{q}$, $p_\mathrm{F}=\sqrt{\mu^2-m^2}$ when $\mu>m$; otherwise, $p_\mathrm{F}=0$. Here, 
$\mu$ and $m$ represent the chemical potential and mass as they appear in the propagator associated with $\vec{q}$. Additionally, the parameter $a$ must satisfy $a>0$. This specific parametrization is chosen because its Jacobian vanishes at the Fermi surface when $a<1$, thereby flattening integrable singularities at the surface (see \cref{sec:multichanneling}). In this work, we have set $a=1/2$.

The above parametrization is general and does not consider any symmetries that the integrand might have. However, in our case, the integrands enjoy rotational symmetry since the vacuum-type diagrams are independent of any external momenta. By exploiting this symmetry, we reduce the dimensionality of the numerical spatial integral from $3l$ to $3l-3$ (or 1 for $l=1$). This reduction is achieved by fixing the polar and azimuthal angles of the first loop momentum such that $\vec{p}_1=(0,0,r_1)$, and fixing the azimuthal angle of the second loop momentum such that $\vec{p}_2=(r_2\sin \theta_2,0,r_2\cos \theta_2)$. This allows for the analytical integration of these trivial angular integrals. The remaining loop momenta are then parametrized with their full angular dependence according to \cref{eq:param}.

\section{Two-loop example for dLTD}
\label{sec:twoloopexample}

Here, we outline some details of the semi-numerical algorithm we use for computing multiloop integrals at finite chemical potential $\mu$. \correctPRD{The example we first consider is the two-loop sunset diagram, which we compare against the analytical bare \MSbar~result \cite{Vuorinen:2003fs} in \cref{eq:table}. We define the bare diagram in Feynman gauge as in \Refe\cite{Vuorinen:2003fs} by}
\begin{equation}
    \raisebox{-0.42\height}{\includegraphics[height=1.4cm]{figures/2loopsunset.pdf}} = -\frac{\gs^2}{2}\nf\mathrm{tr}\left[T^aT^a\right]\int_{\{P_1P_2\}}\frac{\mathrm{tr}\left[\slashed{P}_1\gamma^\mu\slashed{P}_2\gamma_\mu\right]}{P_1^2P_2^2(P_1-P_2)^2},\label{eq:sunsetstart}
\end{equation}
where \correctPRD{$\int_P\equiv(\mathrm{e}^{\gamE}\Lbar^2/(4\pi))^{(4-D)/2}\int\dd^DP/(2\pi)^D$, $\gamE$ is the Euler--Mascheroni constant, and $D$ is the spacetime dimension}. The curly brackets in the integration measure denote fermionic momenta, for which the energy components are shifted by $+i\mu$.

Unlike the UV- and IR-finite combination of the Bugblatter diagrams, the two-loop sunset diagram is superficially UV-divergent \correctPRD{when $D=4$}. These divergences must be subtracted locally at the integrand level for numerical integration \correctPRD{in four dimensions. To achieve this, we employ Bogoliubov's recursive $R$-operation \cite{Bogoliubov:1957gp,Hepp:1966eg,Zimmermann:1969jj}, which subtracts several counterterms $\mathrm{CT}_\Gamma$ from a given Feynman diagram $\Gamma$ to capture all (possibly nested) UV divergences. Within the $R$-operation, we implement local UV renormalization operators $K$, as constructed in \Refe\cite{Capatti:2022tit}, rendering $R(\Gamma)$ UV-finite at the integrand level. Schematically, the UV subtraction for $\Gamma$ has the form}
\correctPRD{
\begin{equation}
    \Gamma = \underbrace{(\Gamma - \mathrm{CT}_\Gamma)}_{D=4} + \underbrace{\mathrm{CT}_\Gamma}_{D\neq 4} \equiv R(\Gamma) + \mathrm{CT}_\Gamma,\label{eq:UVschematic}
\end{equation}
where $R(\Gamma)$ can be computed numerically in $D=4$ dimensions. The remaining counterterms $\mathrm{CT}_\Gamma$ on the RHS of \cref{eq:UVschematic} are computed analytically using dimensional regularization in $D=4-2\varepsilon$ dimensions to obtain the result for the bare diagram $\Gamma$ following the \MSbar~conventions. What makes this approach particularly useful is that $\mathrm{CT}_\Gamma$ is significantly simpler to compute than $\Gamma$ itself because it entails a lower-loop computation.}

In practice, applying the $R$-operation to the sunset diagram yields (for details, see \Refe\cite{Capatti:2022tit})

\begin{widetext}
\begin{equation}
\begin{split}
R\left( \raisebox{-0.42\height}{\includegraphics[height=1.4cm]{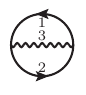}} \right) &= \raisebox{-0.42\height}{\includegraphics[height=1.4cm]{figures/R_op1.pdf}} - \raisebox{-0.42\height}{\includegraphics[height=1.4cm]{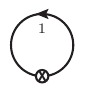}} * K\left( \raisebox{-0.42\height}{\includegraphics[height=1.4cm]{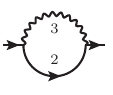}} \right) - \raisebox{-0.42\height}{\includegraphics[height=1.4cm]{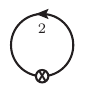}} * K\left( \raisebox{-0.42\height}{\includegraphics[height=1.4cm]{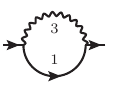}} \right)
- \raisebox{-0.42\height}{\includegraphics[height=1.4cm]{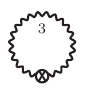}} * K\left( \raisebox{-0.42\height}{\includegraphics[height=1.4cm]{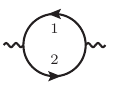}} \right) \\
&- K\left( \raisebox{-0.42\height}{\includegraphics[height=1.4cm]{figures/R_op1.pdf}} \right) + K\left( \raisebox{-0.42\height}{\includegraphics[height=1.4cm]{figures/R_op2.pdf}} * K\left( \raisebox{-0.42\height}{\includegraphics[height=1.4cm]{figures/R_op5.pdf}} \right) \right)
+ K\left( \raisebox{-0.42\height}{\includegraphics[height=1.4cm]{figures/R_op3.pdf}} * K\left( \raisebox{-0.42\height}{\includegraphics[height=1.4cm]{figures/R_op6.pdf}} \right) \right) \\
&+ K\left( \raisebox{-0.42\height}{\includegraphics[height=1.4cm]{figures/R_op4.pdf}} * K\left( \raisebox{-0.42\height}{\includegraphics[height=1.4cm]{figures/R_op7.pdf}} \right) \right),\label{eq:Rop}
\end{split}
\end{equation}
\end{widetext}
where $*$ denotes the contraction of tensor indices, and $K$ is an operator that extracts the UV divergences of the graph it acts on. Specifically, $K(\gamma)$ Taylor expands the integrand corresponding to a graph $\gamma$ in the limit where the loop momenta of $\gamma$ go to infinity, up to the order specified by the superficial degree of divergence (dod) of $\gamma$. Additionally, since thermal effects vanish exponentially in the UV, $K$ sets the chemical potential $\mu$ appearing in $\gamma$ to zero. Consequently, $K(\gamma)$ is a sum of single-scale tensor integrals, where the numerators depend polynomially on the external momenta and masses of $\gamma$, and all the propagators have the same mass $m_\mathrm{UV}$ that is introduced for ensuring the IR-finiteness of the counterterms. As an illustrative example, let us consider one of the $K$-operations appearing in \cref{eq:Rop} explicitly as
\begin{widetext}
\begin{equation}
\begin{split}
K\left(\raisebox{-0.42\height}{\includegraphics[height=1.4cm]{figures/R_op5.pdf}}\right) &= K\left(-\gs^2T^aT^a\int_{\{P_2\}}\frac{\gamma^\mu\slashed{P}_2\gamma_\mu}{P_2^2(P_1-P_2)^2}\right) \\
&= -\gs^2T^aT^a\left\{\int_{P_2}\frac{\gamma^\mu\slashed{P}_2\gamma_\mu}{\left(P_2^2+m_\mathrm{UV}^2\right)^2} + \int_{P_2}\frac{2P_1\cdot P_2\gamma^\mu\slashed{P}_2\gamma_\mu}{\left(P_2^2+m_\mathrm{UV}^2\right)^3}\right\}, \label{eq:Kexp}
\end{split}
\end{equation}
\end{widetext}
where the expansion is carried out to the next-to-leading order dictated by the graph's superficial dod of one. The expanded tensor integral in \cref{eq:Kexp} is then contracted with the remaining graph, resulting in
\begin{widetext}    
\begin{equation}
\raisebox{-0.42\height}{\includegraphics[height=1.4cm]{figures/R_op2.pdf}} * K\left(\raisebox{-0.42\height}{\includegraphics[height=1.4cm]{figures/R_op5.pdf}}\right) = -\frac{\gs^2}{2}\nf\mathrm{tr}\left[T^aT^a\right]\left\{\int_{\{P_1\}P_2}\frac{\mathrm{tr}\left[\slashed{P}_1\gamma^\mu\slashed{P}_2\gamma_\mu\right]}{P_1^2\left(P_2^2+m_\mathrm{UV}^2\right)^2} + \int_{\{P_1\}P_2}\frac{2P_1\cdot P_2\mathrm{tr}\left[\slashed{P}_1\gamma^\mu\slashed{P}_2\gamma_\mu\right]}{P_1^2\left(P_2^2+m_\mathrm{UV}^2\right)^3}\right\}. \label{eq:Kexp2}
\end{equation}
\end{widetext}
The remaining counterterms in the first row of \cref{eq:Rop} are computed analogously. For the second and third rows, the $K$-operation acts on vacuum-type graphs with a dod of four. In these cases, we simply set $\mu=0$ without performing any expansions.

With the sunset diagram now rendered locally finite through the $R$-operation, we proceed to the analytical integration of the energy integrals using the dLTD formula in \cref{eq:Iresult}. The form of the counterterms (see \cref{eq:Kexp2}) is compatible with the starting point of dLTD in \cref{eq:Iintegral}, allowing us to input \cref{eq:Rop} directly into our algorithm that computes the energy integrals. The higher-order residues introduced by degenerate propagators appearing in the counterterms are accessed by taking derivatives with respect to the on-shell energies of these propagators. Next, we provide an example computation of the energy integrals in the first term on the right-hand side of \cref{eq:Rop}, which corresponds to the bare sunset diagram.


Comparing \cref{eq:sunsetstart} and \cref{eq:Iintegral}, we obtain $l=2$, $n=3$, $\mathbf{S}_1=(1,0,1)$, $\mathbf{S}_2=(0,1,-1)$, $\boldsymbol{\varphi}=(1,1,0)$, and $\mathbf{E}=(|\vec{p}_1|,|\vec{p}_2|,|\vec{p}_1-\vec{p}_2|)$. To perform the $\boldsymbol{\tau}$-integrals in \cref{eq:Iresult}, corresponding to Laplace transforms of nonsimplicial convex cones, we use the diagrammatic algorithm introduced in \Refe\cite{Capatti:2022mly}. Specifically, the sum over $\boldsymbol{\sigma}$ is interpreted as a sum over directed graphs, where only the acyclic ones contribute. For each of these acyclic graphs, a series of edge-contraction operations is performed, ultimately yielding a closed-form expression for the corresponding Laplace transform. For instance, one of the nonvanishing terms in the current example, with $\boldsymbol{\rho}=(1,-1,1)$ and $\boldsymbol{\sigma}=(1,-1,-1)$, gives
\begin{equation}   
\begin{split}    
&\int_{\mathbb{R}_+^3}\dd\tau_1\dd\tau_2\dd\tau_3\,\mathrm{e}^{-\tau_1E_1+\tau_2E_2-\tau_3E_3}\delta(\tau_1-\tau_3)\delta(-\tau_2+\tau_3)\\
& =\frac{1}{E_1-E_2+E_3}.
\end{split}
\end{equation}
Performing all the $\boldsymbol{\tau}$-integrals in \cref{eq:Iresult} for the bare sunset results in
\begin{widetext}    
\begin{equation}
\begin{split}
    \raisebox{-0.42\height}{\includegraphics[height=1.4cm]{figures/2loopsunset.pdf}} &= 2\gs^2\nf(\nc^2-1) \int\frac{\dd^3\vec{p}_1}{(2\pi)^3}\frac{\dd^3\vec{p}_2}{(2\pi)^3}\Biggl\{\frac{(E_1E_2+\vec{p}_1\cdot\vec{p}_2)\Theta(E_1-\mu)}{2E_12E_22E_3(E_1+E_2+E_3)}-\frac{(-E_1E_2+\vec{p}_1\cdot\vec{p}_2)\Theta(E_1-\mu)\Theta(-E_2+\mu)}{2E_12E_22E_3(E_1-E_2+E_3)} \\
    &+\frac{(E_1E_2+\vec{p}_1\cdot\vec{p}_2)\Theta(E_2-\mu)}{2E_12E_22E_3(E_1+E_2+E_3)}-\frac{(-E_1E_2+\vec{p}_1\cdot\vec{p}_2)\Theta(-E_1+\mu)\Theta(E_2-\mu)}{2E_12E_22E_3(-E_1+E_2+E_3)}\Biggr\},\label{eq:sunset3d}
\end{split}
\end{equation}
\end{widetext}
where only four distinct terms contribute. Note that most of the $2^{2n}$ ($n=3$) terms vanish because they correspond to cyclic graphs or include a vanishing step function. By applying the same procedure to the seven counterterms to perform the energy integrals, we are now ready to proceed to the numerical evaluation of the spatial integrals in \cref{eq:Rop}.

The remaining three-dimensional spatial integrals in the UV-subtracted sunset diagram contain integrable singularities that hinder numerical integration. These singularities are flattened using the multichanneling procedure described in \cref{sec:multichanneling}. For the sunset diagram, there are three channels in total, specified by the loop momentum bases $\mathcal{B}=\{\{1,2\},\{1,3\},\{2,3\}\}$. Each channel is then integrated separately using \textsc{Vegas5.6}. With roughly $10^8$ samples in total, the sum of the results for the three channels is
\begin{equation}
R\left( \raisebox{-0.42\height}{\includegraphics[height=1.4cm]{figures/2loopsunset.pdf}} \right) = -0.00128338(23)\cdot\mu^4,\label{eq:sunsetres}
\end{equation}
where $\gs=\nf=1$, $\nc=3$, and $m_\mathrm{UV}=\mu$. \correctPRD{To obtain the \MSbar~result for the bare sunset diagram in \cref{eq:sunsetstart}, we must compute the subtracted counterterms analytically using dimensional regularization and then add them back to \cref{eq:sunsetres}, as shown schematically in \cref{eq:UVschematic}. However, the counterterms in \cref{eq:Rop} are proportional to scaleless integrals, which evaluate to zero in dimensional regularization. Therefore, \cref{eq:sunsetres} directly gives the bare \MSbar~result that can be compared with the known value in \Refe\cite{Vuorinen:2003fs}. The comparison shows an agreement within a relative error of $0.02\%$.}

In \cref{eq:table}, we also present \correctPRD{the \MSbar~result for the finite $O(\varepsilon^0)$ part of} the three-loop QCD ``Mercedes" diagram \correctPRD{in Feynman gauge}. The algorithmic steps in the dLTD evaluation of this diagram follow those used in the two-loop sunset case discussed above. Compared to the two-loop case, the UV-subtracted Mercedes contains 21 counterterms, the bare diagram consists of 72 terms after performing the zero-component integrals (cf. \cref{eq:sunset3d}), and the multichannel procedure involves 16 channels. With roughly $10^9$ Monte-Carlo samples we obtain the following result for the UV-subtracted Mercedes:
\begin{equation}
R\left( \raisebox{-0.42\height}{\includegraphics[height=1.4cm]{figures/3loop5.pdf}} \right) = -1.4239(9)\cdot 10^{-4}\cdot\mu^4,\label{eq:mercedesres}
\end{equation}
where $\gs=\nf=1$, $\nc=3$, and $m_\mathrm{UV}=\mu$. \correctPRD{Once again, we add the analytically computed counterterms back to \cref{eq:mercedesres} to obtain the \MSbar~result for the bare Mercedes diagram. Unlike the two-loop case, the counterterms for the Mercedes do not vanish in dimensional regularization. Instead, by computing these counterterms in $4-2\varepsilon$ dimensions and adding them to \cref{eq:mercedesres} we obtain}
\begin{equation}
\begin{split}
\raisebox{-0.42\height}{\includegraphics[height=1.4cm]{figures/3loop5.pdf}} &= \biggl(-7.3136\cdot 10^{-5}\cdot\frac{1}{\varepsilon} \\
&\quad -0.00025654(9)+O(\varepsilon)\biggr)\cdot\mu^4,\label{eq:mercedesres2}
\end{split}
\end{equation}
where we have set $\Lbar=\mu$. \correctPRD{The $O(\varepsilon^0)$ coefficient of} this result agrees with the known value \cite{Vuorinen:2003fs} within a relative error of $0.03\%$.

\section{Regularization of non-local IR singularities in \cref{b1}}
\label{sec:gluonmass}

\begin{figure}[H]
    \centering
    \vspace{2em}
    \includegraphics[width=0.48\textwidth]{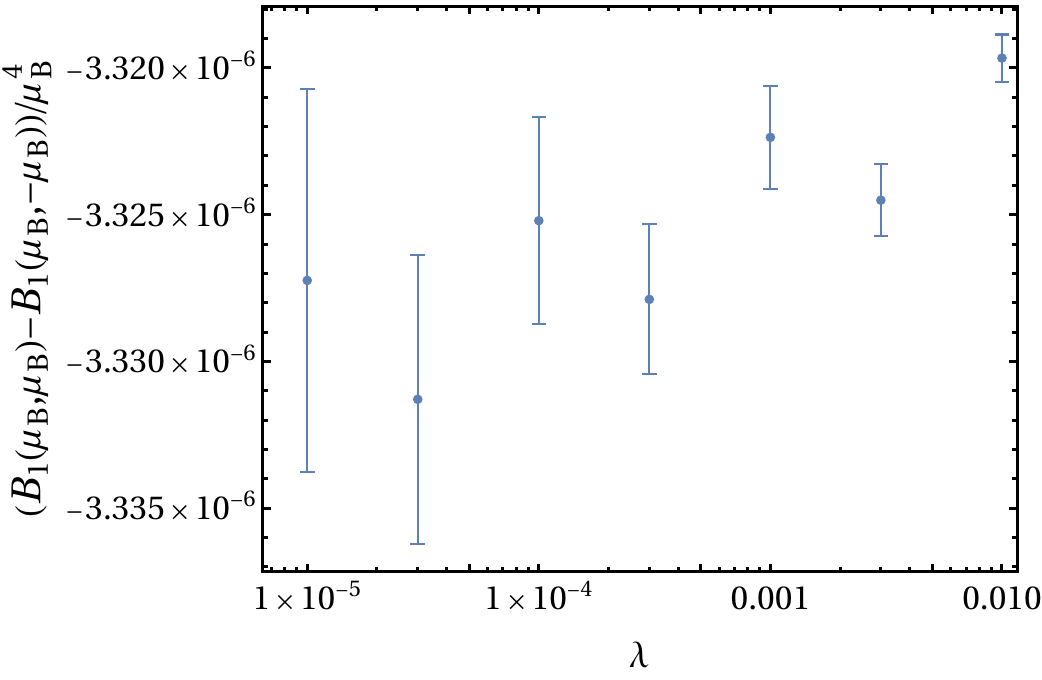}
    \caption{Results for the numerical integration of \cref{b1} using seven different values for the fictitious gluon mass parameter $m_g=\lambda\muB$.}
    \label{fig:lambdaplot}
\end{figure}

The cancellation of IR divergences in \cref{b1} is non-local, necessitating a regularization procedure for numerical evaluation. To address this, we introduce a small but nonzero mass in the gluonic on-shell energies as an IR regulator. Since the combination of Bugblatter diagrams we compute is IR safe after integration, the effect of the IR regulator should be insignificant, provided its numerical value is much smaller than the dominant scale of the integral, namely the baryon chemical potential $\muB$. We verify the insensitivity by evaluating the integral with several different small values of the gluon mass $m_g$.
Specifically, we define $m_g=\lambda\muB$ and compute the integral in \cref{b1} using seven values of $\lambda$ within the range $[10^{-5},10^{-2}]$. The results are shown in \cref{fig:lambdaplot}. In the figure, we observe that all results are approximately equal within the error bars, except for the one with the largest IR regulator, $\lambda=10^{-2}$. Therefore, as long as the fictitious gluon mass remains well below one percent of $\muB$, the result is unaffected. However, for IR regulator values with $\lambda$ less than $10^{-5}$, the convergence rate of the numerical Monte Carlo integration significantly slows down. For the final result reported in the main text, we choose $\lambda=10^{-5}$ as the most conservative option.

\bibliography{references}

\end{document}